# Memory in strain-tuned insulator-metal-insulator sequence of transitions after photoexcitation in the Mott material V$_2$O$_3$

O. Yu. Gorobtsov[1]*, Y. Kalcheim[2,3], Z. Shao[1], A. Shabalin[2], N. Hua[2], D. Weinstock[1]†, R. Bouck[1], M. Seaberg[4], D. Zhu[4], O. G. Shpyrko[2], I. K. Schuller[2], and A. Singer[1]*

[1] Department of Materials Science and Engineering, Cornell University; Ithaca, 14853, USA.

[2] Department of Physics, University of California San Diego; La Jolla, 92093, United States.

[3] Department of Material Science and Engineering, Technion-Israel Institute of Technology; Haifa, 32000, Israel.

[4] SLAC National Accelerator Laboratory; Menlo Park, 94025, USA.

*Corresponding authors. Email: gorobtsov@cornell.edu, asinger@cornell.edu

†Author deceased

**ABSTRACT**

Memory effects during metal-insulator transitions in quantum materials reveal complex physics and potential for novel electronics mimicking biological neural systems. Nonetheless, understanding of memory and nonlinearity in sequential non-equilibrium transitions remains elusive as the full chain of transitions can involve features lasting anywhere from femtoseconds to microseconds. Here, we extend time-resolved x-ray Bragg diffraction to the dynamic range of timescales spanning 9 orders of magnitude to fully trace the pathways of photoexcited insulator-metal transition and the following relaxation through non-equilibrium metal-insulator transitions in epitaxial films of V$_2$O$_3$, a promising Mott material. We find 5 orders of magnitude variation in metal-insulator transition time, from nanoseconds to hundreds of microseconds, depending on pre-excitation phase state. We provide a theoretical explanation and simulations based on strain feedback to domain nucleation. The lingering transition is stretched in time by memory of spatial and energy heterogeneity and, contrary to known memory effects in vanadium oxides commonly described by power laws, follows an extremely (factor below 0.2) stretched exponential. The induced dramatic slowdown in the light-driven highly correlated system signifies unusually high heterogeneity of transition barriers similar to biological systems, and demonstrates importance of non-local correlations of structure in evolution of transitional phases in quantum materials.

# MAIN

Energy barriers in material phase transitions constrain property fluctuations, providing the physical basis for memory storage[1,2]. In quantum materials with electronic phase transitions, these kinetic constraints enable the combination of variable electronic resistivity and memory, laying the foundation for memristive and neuromorphic functionality in advanced computing systems. Mott insulators stand as promising candidates for applications in computing and photonics due to a reversible switching between well-defined insulating and metallic states[3-5]. In a prototypical Mott system vanadium sesquioxide ($V_2O_3$)[6], the switching occurs within picoseconds after optical excitation[7,8] and within nanoseconds after electrical pulses[3], demonstrating nonlinear response and remarkable transition speed promising novel computing device applications[9-12]. Still, significant challenges for application persist: the photoinduced metallic state is typically short-lived; and the coupling between charge carriers and structure remains poorly understood, largely due to the lack of techniques that simultaneously resolve structural dynamics across wide temporal ranges.

Here, we demonstrate a dramatic adjustment in the full transition time by a factor of $10^5$ dependent on the structural phase proportion before the photoexcitation. We track the transition from an insulating Mott state into a photoexcited metallic state and the relaxation back into the insulating state over timescales from 100 femtoseconds to microseconds using high dynamic range time-resolved X-ray diffraction at a Free Electron Laser (FEL). We effect the slowdown of relaxation by thermally manipulating the initial state of the epitaxially grown thin film: we achieve a 3-nanosecond relaxation time when launching the photo-induced transition from a fully insulating state versus a 100-microsecond relaxation when launching from a state with coexisting metallic and insulating domains. The kinetically inhibited relaxation follows Kohlrausch-Williams-Watts (KWW) stretched exponential behavior[13,14] characteristic of systems with a broad distribution of activation energies[15]. We explain inhibited relaxation through a microscopic model, accompanied by numerical simulations, where self-limiting kinetics modulate the energy barrier distribution via a feedback loop between the strain and the strain-dependent transition temperature. Our model, incorporating simultaneous nucleation and growth of phase domains under varying strain conditions, provides a framework to understand and potentially control the complex relaxation pathways in quantum materials with discontinuous phase transitions. Our results enable potential design of materials taking advantage of transient memory states and non-linear, hysteretic, and reproducible responses with heterogeneous energy barriers that parallel "anomalous" processes in strongly correlated biological neural systems[16,17].

We synthesized 100 nm $V_2O_3$ thin films on a sapphire substrate (M-plane: $(10\bar{1}0)$) with an approach similar to references[3,18]. The films are metallic above and insulating below the transition temperature of approximately 170 K, with a hysteresis of 7 K[18] and phase coexistence in a similar temperature range. In a quasi-static temperature-driven transition, electronic transition is coincident with the structural one from the lower symmetry monoclinic insulating phase to the higher symmetry rhombohedral metallic phase. By tuning the temperature, we prepared ground states with different phase ratios and optically excited the films with 50 fs short laser pulses at an energy of 1.54 eV, which is above the bandgap of 0.6 eV in bulk $V_2O_3$, ensuring efficient carrier excitation. We calibrated the fluence of the optical laser to achieve full transformation into the high-temperature phase in 100 ps after the photoexcitation without excess heating (see Materials and Methods, fig. S1). We used the reciprocal space around the specular 300 X-ray diffraction peak from the rhombohedral structure as a probe of the structural response during the transition. We optimized experimental geometry so that the Ewald sphere cuts through diffraction peaks from different structural phases without the need for rocking the sample.

Figure 1A displays the 2D detector image of the diffraction intensity in and around the Bragg peak position when the film is entirely in the low-temperature monoclinic ground state[18]. Photoexcitation drives the system into the high-temperature rhombohedral phase, evidenced by the changes in diffraction data at 200 ps delay (Fig. 1B): low-temperature monoclinic phase peak disappears and inthestead a peak arises at a different position (inside blue rectangle) corresponding to high-temperature rhombohedral phase. The weak Bragg peak at the lower q value is due to a high-temperature rhombohedral, paramagnetic insulating phase[18]. When the system relaxed into the ground state, the diffraction image is indistinguishable from the one from the ground state (Fig. 1C). We quantified the time-dependent phase ratio by integrating the diffraction signal over the region highlighted in Figs. 1A-C and normalizing the total integrated intensity to 1 (Fig. 1D). During the first stage – the photoexcited insulator-to-metal transition – two distinct timescales emerge. First, a few picoseconds structural response (kinks marked by arrows Fig. 1D, inset) previously[7] connected to decrease of monoclinic distortion after photoexcitation towards a more symmetric rhombohedral structure. Second, a separate diffraction peak emerging with a characteristic time of 50 ps (Fig. 1D) and indicating a transition into the high-temperature phase with a different lattice constant. When the initial state comprises both phases (0.65 for the low-temperature monoclinic and 0.35 for the high-temperature rhombohedral, Fig. 1D, black crosses), the characteristic times for reduction in monoclinic distortion and the transition to high-temperature phase remain similar to those we observed for the excitation from the pure low-temperature ground state (Fig. 1D, red circles).

After the system fully transitions into the high-temperature rhombohderal structure, the second stage – relaxation back into the ground state – starts. During this second stage, a dramatic difference emerges between the processes with different initial state (Fig. 1E). The relaxation into the low-temperature ground state has a characteristic time $\tau_{r,0} \equiv \tau_r(\Phi_{HT,g} = 0) = \sqrt{\langle t^2 \rangle} = 4\ ns$ (Fig. 1E, inset shows semilogarithmic plot). In contrast, when the initial state comprises both phases, the characteristic time of the relaxation increases by multiple orders of magnitude. The relaxation is incomplete after 2 microseconds (Fig. 1E, maximum measured time delay).

Notably, the relaxation process is broadly inconsistent with exponential decay(s) nor does it follow a power law (Fig. 1F shows best fits in logarithmic scale). Instead, we describe the evolution of the instantaneous HT phase ratio, $\Phi_{HT}(t)$, via a Kohlrausch-Williams-Watts (KWW) (13,14) function

$$\Phi_{HT}(t) = \Phi_{HT,g} + A \cdot \exp[-(t/\tau_K)^\beta], \quad (1)$$

where $\Phi_{HT,g}$ is the phase ratio in the ground state (lowest energy state for the system in equilibrium conditions), $A$ and $\tau_K$ are constants, and $\beta$ is the stretching parameter. To isolate the long-term process and determine its parameters, we limit the fitting to the heavy tail of the relaxation curve, after the excitation into the high-temperature state is complete. The stretched exponential model accurately reproduces the experimental relaxation dynamics over more than 2 orders of magnitude in time and yields $\beta = 0.15$ (Fig. 1F) with $A = 10$ and $\tau_K = 10\ ps$. The constant $\tau_K$ gravely mischaracterizes relaxation time if the aim is to compare with time constants typically obtained from fitting functions such as a simple exponential[19]. The standard deviation (analogous with the universal approach to distribution widths) calculated from the fit gives $\tau_{r,0.35} = \sqrt{\langle t^2 \rangle}$=150 µs. This value is $10^5$ times larger than $\tau_{r,0} = 4\ ns$ characteristic time of the relaxation when the ground state is purely monoclinic. The dramatic dependence of relaxation time on the ground state phase ratio is the key finding of this study. To confirm the effect, we performed the measurements on different standard substrate orientations. The timescales remain consistent across films grown on all measured substrate orientations A, R, and M (Fig. 1E), indicating this behavior is intrinsic to $V_2O_3$ thin films (Fig. 1G).

The extraordinarily long characteristic relaxation time suggests significantly kinetically inhibited transition. Furthermore, Kohlrausch-Williams-Watts behavior typically emerges in spatially heterogeneous materials with anti-cooperative dynamics, where anti-cooperative behavior describes how transformation becomes inhibited in areas adjacent to transformed regions. While a quantum material possesses many degrees of freedom that could display anti-cooperative behavior, the observed timescales (microseconds) exclude electronic interactions and thermalization as an explanation, as these are typically much faster in thin films (picosecond and nanosecond timescale[7,20]). Thermal explanation is further excluded as the amount of heat to transport from the film to the substrate is even lower than for the fast transition, limiting the timescale of the heat transport to below 10 ns.

We propose, therefore, that this anti-cooperative behavior stems from interaction between strain and local structural order, recently identified as a key parameter in the photo-induced transformation in $VO_2$ (21). As regions transform from high-temperature rhombohedral phase to the larger-volume (22) low-temperature monoclinic phase, the film experiences increasing in-plane compression due to substrate clamping and out-of-plane expansion due to Poisson effect. Figures 2A, 2B (supported by fig. S2) confirm this prediction, showing out-of-plane expansion during transformation in the low-temperature monoclinic peak with a progressive decrease during relaxation.

The hypothesis of anti-cooperative behavior must lead to smaller general size of domains at a certain phase ratio due to limits on domain growth, leading to higher surface of interfaces between the domains of structural phases. We performed analysis of the relative interface proportion through the transition based on microstructure-caused changes in the shape of the Bragg peaks[23]. To quantify the density of interfaces between the low-temperature and high-temperature phases—which reflects the number of discrete low-temperature domains—we monitored the diffuse scattering adjacent to the main Bragg peak (see Figs. 1 and 3), oriented perpendicular to q. This scattering tail is present only at intermediate temperatures (see Fig. 1a; marked with blue square in Fig. 2C, D) corresponds to a structure existing on the interface between domains of rhombohedral and monoclinic structural phases[18]. When the initial state is fully monoclinic, the intermediate-temperature peak transiently arises during the metal-insulator relaxation, reaching the maximum phase fraction at 3 ns (Fig. 2C, E). The maximum scattering volume of the interface is smaller than in quasi-static transition, as visible from the initial signal in Fig. 2F. The scattering volume of the intermediate state is significantly higher during relaxation into the heterogeneous metal-insulator state, consistent with the expectation of smaller domains and therefore increased spatial heteregeneity.

We develop a qualitative model based on the Johnson–Mehl–Avrami–Kolmogorov (JMAK) approach[24] to describe the stretched exponential relaxation. The classical JMAK model is based on the assumption that the change of the spatially averaged phase ratio at time $t$ of the high-temperature phase $\Phi_{HT}$ is given by $d\Phi_{HT} = -\Phi_{HT}\dot{N}G(t-\tau)dt$, where $\dot{N}$ is the nucleation rate, typically assumed constant, and $G(t-\tau)$ is the increase in the phase volume of a single domain due to the growth of the nucleus formed at time $\tau$. In the classical model, the system always fully transitions into the end state, because $\dot{N}$ and $G(t-\tau)$ remain nonzero until the transition is complete. To limit the transition and reproduce the experimentally measured phase coexistence after relaxation, we assume that (1) the nucleation rate depends on the phase ratio $\dot{N}(\Phi_{HT})$ and approaches 0: $\dot{N}(\Phi_{HT,g}) = 0$ (as there is no change in ground state) and (2) the domain growth stops when the domain size reaches a value of $R_{max}$ (the simplest sufficient assumption). We assume that the growth is fast, and the nucleus reaches the maximum size rapidly. These modifications result in a change of the phase ratio

$$d\Phi_{HT} = -\Phi_{HT}\dot{N}R_{max}dt. \quad (2)$$

Solving this equation with experimentally measured $\Phi_{HT}(t)$ allows us to calculate the nucleation rate dependence on $\Phi$ from experimental data (Fig. 3A). The observed nucleation rate smoothly decreases from initial values, satisfying a simple sanity check, and continuously approaches 0 as the phase ratio approaches the final values, satisfying the boundary condition. Continuous approach of the nucleation rate to 0 (Fig. 3A, Materials and Methods) as a function of the driving force, directly dependent on phase ratio in this model, also confirms the physicality of our model (compare to (25), Fig. 19.11).

We introduce a further spatially resolved model to simulate the evolution of the microstructure. We describe the dynamics of the local order parameter $\varphi_{HT}(x,t)$ via an Allen-Cahn type equation[25], introducing state-dependent energy to simulate the slowdown we observe experimentally:

$$\frac{\partial \varphi_{HT}(x,t)}{\partial t} = M(\varepsilon \Delta_x \varphi_{HT}(x,t) - E'(\varphi_{HT})), \quad (3)$$

where $M$ is the mobility and coefficient $\varepsilon$ characterizes the strength of the term arising due to spatial inhomogeneity of the phase distribution. We simulate the discontinuous transformation by introducing the homogeneous Landau-Ginzburg free energy of a two-phase system, $E(\varphi_{HT})$, as a double well potential with minima at 0 and 1 (representing the two phases at $\varphi_{HT} = 0$ and $\varphi_{HT} = 1$ separated by an energy barrier, signifying first order transition[26]) and a saddle point at $\varphi_c$ (Fig. 3B). The energy gradient is then given by $E'(\varphi_{HT}) = dE(\varphi_{HT})/d\varphi_{HT} = 4a\varphi_{HT}(\varphi_{HT} - \varphi_c)(\varphi_{HT} - 1)$, where $a$ is a constant. To simulate the change in the energy barrier and the energy difference between the two stable phases throughout the phase transformation we define the position of the saddle point to depend on the mean field phase ratio $\Phi_{HT}$ via $\varphi_c(\Phi_{HT}) = 0.5\left(1 + \frac{|\Phi_{HT}-\Phi_{HT,g}|^2}{(1-\Phi_{HT,g})^2}\right)$. Here, the power of 2 is suggested by the series of simulations as the closest integer value to reproduce the stretched exponential, implying a second power of the phase ratio in the Ginzburg-Landau free energy of the system[26]. At the beginning of the relaxation, $\Phi_{HT} = 1$ and the energy barrier is absent as $\varphi_c = 1$. As $\Phi_{HT}$ approaches $\Phi_{HT,g}$, the energy barrier increases, decreasing the transformation rate, and $\varphi_c = 0.5$ so the kinetic rate balance stabilizes the average phase ratio at $\Phi_{HT} = \Phi_{HT,g}$ (Fig. 3B). We numerically solve equation (3) (see Materials and Methods), performing a time-evolving simulation. Fig. 3C shows the evolution of $\Phi_{HT}(t)$ that closely follows a stretched exponential behavior with $\beta = 0.11$. Inset domain maps show increasing area of the nucleated phase as transition progresses. Inset plot shows interface ratio evolution as a function of time that demonstrates similar behavior to experimental – a growth of interface phase ratio with following decrease.

Our high dynamic range time-resolved X-ray diffraction experiments on photoexcited $V_2O_3$ reveal a complex transition process with features spanning 9 orders of magnitude in time, culminating in a remarkably slow (~100 µs) relaxation with memory that can be induced through the ground state manipulation. The relaxation is slowed down by 5 orders of magnitude. The wide range of timescales of our method let us identify this relaxation as a Kohlrausch-Williams-Watts (stretched exponential) process with excellent agreement between the fit and the experimental data allowing to exclude power law or simple exponential as description.

KWW dynamics often arise due to anti-cooperative behavior in disorganized and self-organized systems ranging from glasses and proteins to human brain[16,27,28]. While in photoexcited thin film quantum materials, changes in timescales such as critical slowing down[29] in non-equilibrium conditions have been observed due to electronic[30,31] or topological[32] processes, the timescales involved are generally on picosecond timescales, including in $VO_2$[20]. The extremely low nucleation rate derivative near the ground state phase ratio (see inset in Fig. 3A and fig. S3) still presents a critical slowing down as the system approaches equilibrium, but on an entirely different timescale. In the context of a spatially heterogeneous system such as a $V_2O_3$ film with coexistent domains of different structural phases, we instead propose a microstructure-based, spatially anti-cooperative mechanism based on strain, which can be illustrated on the equilibrium phase diagram of $V_2O_3$ (Fig. 3D). As the excited rhombohedral structure starts to transition into monoclinic structure after thermalization, the growth of the monoclinic structure with a larger volume results in effective compression and shifts the system towards the phase boundary (akin to high pressure), reducing the thermodynamic driving force and increasing the transition energy barrier. This self-limiting process can be interpreted as leading to a broad energy barrier distribution, which can be directly extracted[15] from experimental results (Fig. 3E).

Notably, we find an exceptionally low stretching factor $\beta=0.15$ across 3 different film orientations (M, R, A substrate cuts), signifying an extreme heterogeneity in dynamics and particularly broad energy barrier distribution. Consider that in most glassy systems, amorphous semiconductors, proteins, superconductors, and other widely known systems typical values are $\beta \sim 0.3\text{-}0.7$[16,27,28,33]; models commonly used in some systems, such as percolation model in glasses or effective dimensionality approaches, in fact forbid values below $0.3$[33,34]. Lower values tend to be found in "jammed" systems rich with metastable states, such as catalytic activity of proteins with complex conformational dynamics[35]. A relevant consideration is the difference between how "serial" (possible energy state evolve with time) or "parallel" (whole barrier distribution exists in every moment of time) the process is.

To rationalize the experimentally measured KWW behavior with an extremely low stretching factor, used two complementary approaches: a modified Johnson–Mehl–Avrami–Kolmogorov model, a "serial" model with changing domain nucleation rates and limited domain size, and an Allen-Cahn type phase field simulation with self-limiting through changing energy landscape. Both models not only let us reproduce the stretched exponential behavior with low $\beta$, but in the case of Allen-Cahn simulation also reproduce prolonged strain relaxation and interface volume change consistent with what we found experimentally. We thus suggest the anti-cooperative spatial mechanism through non-local strain as the physical basis for the observed memory effect. Future research should also explore micro- and nanostructure measured directly or indirectly in time-resolved experiments to clairfy potential interfacial and non-equilibrium scarring[36] (Supplementary Text) and domain maps.

Our findings highlight the importance of non-local structural interactions in non-equilibrium phase transition dynamics. The ability to resolve these dynamics across picosecond to microsecond timescales, enabled by XFEL-based diffraction, was essential to determine the nature of relaxation. Based on existing coupling between electronic and structural properties in Mott insulators, the structural memory intrinsic to the photoexcited phase transition raises questions about accompanying changes in spatially resolved measurements of the pattern changes on the timescales we found, either through imaging or speckle pattern interferometry[37], as well as electronic properties measurements. Furthermore, effects of other external interactions (strain, current) on the pattern memory must be investigated in the future to expand control over the mechanism. One of the examples would be the effect of repeated stimulation to further transition from "short" to "long" term memory[38], suggesting a route towards tunable memory based not on residual structure, but on emergent barrier landscapes shaped by prior photoexcitation – an application of "properties on demand"[39] in a quantum material. We expect that our findings are generalizable to a wide range of systems with volume change and substrate clamping due to the presence of the effect over different substrate orientations. Our results also suggest how knowledge of the phase diagram in other materials can imply conditions under which similar memory effects can be found (for example, in $VO_2$[40]), predicting non-equilibrium behavior based on equilibrium phase diagram. The knowledge of and ability to manipulate non-equilibrium metal-insulator phase transitions with emergent memory through landscapes unusually rich with metastable states provide, therefore, a physical basis for advances in memory technology and for mimicking biological neural systems with similar KWW dynamics.

## METHODS

### Sample Preparation

100 nm V2O3 films were grown on sapphire (Al2O3) substrates by RF magnetron sputtering as described previously in (41,42).

### Experimental parameters and calibration

The pump-probe experiment was carried out at the XPP instrument of the LCLS with an x-ray photon energy of 8.9 keV, selected by the (111) diffraction of a diamond crystal monochromator. The x-ray polarization is horizontal, and the scattering geometry is vertical. X-ray diffraction in the vicinity of the specular out-of-plane Bragg peaks from each pulse was recorded by an area detector (CS140k) with a repetition rate of 120 Hz. The film was cooled with a cryojet. About 100 pulses were recorded per time delay. The sample was excited by optical (800 nm, 40-fs), p-polarized laser pulses propagating nearly collinear with the x-ray pulses. The incident fluence on the sample was controlled by the angle of an optical waveplate between polarizers and measured with a photodiode. The final temporal resolution was estimated to be 80 fs. The fluence of the optical laser was calibrated for different sample cuts and pre-excitation conditions by measuring the dependence of post-excitation maximum high-temperature rhombohedral diffraction peak intensities as a function of fluence. Such dependence shows saturation at a fluence where full transition is achieved; any additional laser power only overexcites the sample. The fluence was therefore calibrated effectively so that agnostic to the pre-excitation condition, the after-excitation full transition state is always reached.

**Estimates of the characteristic relaxation/excitation times**

To uniformly compare the timescales between processes defined by different non-exponential behaviors, we have used the standard deviation approach to calculate the characteristic times. Namely, a relaxation or excitation curve is centered on the starting point of the process, normalized by the area under the curve, mirrored across the center to produce a distribution, and the standard deviation is calculated:

$$\sigma = \left( \mathbb{E}[x(t)^2] / \int_{-\infty}^{\infty} x(t) dt \right)^{1/2}$$

**Modified Johnson–Mehl–Avrami–Kolmogorov model**

In the classic derivation of the Johnson–Mehl–Avrami–Kolmogorov equation, a constant specific (by volume) nucleation rate $\dot{N}$ leads to a continuous generation of domains of the nucleating phase within the remaining volumes of the original phase. Every domain then continues to grow linearly with time in every dimension (2 in the case of a thin film), with a volume $G(t - \tau)$ growing polynomially, until the whole material is transformed into the new phase:

$$d\Phi_{HT} = -\Phi_{HT} \dot{N} G(t - \tau) dt.$$

It is easy to see that this dependence leads to a faster-than-exponential growth, as the constant $\dot{N}$ would already produce an exponential; indeed, in the classic JMAK

$$\Phi_{HT} = 1 - e^{-Kt^n},$$

where n is the dimensionality. Therefore, to produce both stabilization of the phase ratio and a slower-than-exponential growth, both the domain growth must be limited and the nucleation rate must decrease as the system approaches equilibrium. Pointedly, domain growth speed can be assumed to be much faster than the typical KWW process we observe, based on the behavior of the system when $\Phi_{HT,g} = 0$; and the limit on domain size must exist, otherwise the entire system will be transformed into the new phase. Therefore, in the first approximation, we can replace the domain growth term with an assumption of a size-limited ($R_{max}$) domain being produced "instantaneously" (in KWW scale) after nucleation. The behavior of the relaxation process is then entirely contained in the specific nucleation rate, and

$$d\Phi_{HT} = -\Phi_{HT} \dot{N} R_{max} dt.$$

Provided with experimental knowledge of the time-dependent phase ratio, we then directly extract the specific nucleation rate as a function of time, as shown in the main text. Note that this effectively translates the KWW behavior into the question of nucleation rate slowdown, which narrows the physical possibilities behind the mechanism.

**Allen-Cahn type simulation**

Simulation of domain evolution requires a more developed model. The existence of two phases and evolving dynamical balance between them suggests a double-well local energy potential with a changing barrier and relative well depth. For the potential shape described in the main text, the process stops as $\varphi_c$ approaches 0.5. Note that if $\Phi_{HT}$ is close to zero, the model is not valid (such as when $\Phi_{HT,g} \sim 0$), as the underlying assumptions are no longer fulfilled. A different model would be required, based on an entirely separate set of measurements which is beyound the scope of this paper.
Simulation parameters: 1000x1000 grid dimensions; calculation parameters in arbitrary units 0.1 grid spacing, 0.01 numerical time step, 5000 iterations.

**Zeroth-order approximation behind the stretched exponential mathematical description**

We can describe the relaxation process phenomenologically by constructing a model relying on the following observations: 1) the relaxation process is *non-exponential* (Fig. 1,d) and the behavior in any given time interval depends on the delay, i.e., the instantaneous dynamics depend on the preceding pathway of the phase transformation (memory effect); 2) the relaxation process does *not* follow a power law (Fig. 1,d); 3) the relaxation process depends on the ground state phase ratio; 4) the relaxation with memory is much slower than without. Due to (4), we can effectively separate the relaxation process into two: 1) a "fast" (nanoseconds) process responsible for relaxation without memory, which we assume to be exponential: $\dot{\Phi} \sim (\Phi - \Phi_b)$, where $\Phi$ is the instantaneous HT phase ratio and $\Phi_b$ is the base phase ratio – equal to the final phase ratio after relaxation if there is no memory effect and the slower process is absent; and 2) a much slower (microseconds) process driving the changes in that base phase ratio and responsible for the memory: $\Phi_b = \Phi_g + \varphi(\Phi_b, t)$, where $\Phi_g$ is the final ground state phase ratio. We can assume in the first approximation that $\dot{\varphi}(\Phi_b, t) = (\Phi_b - \Phi_g) f(t)$, and then the qualitative differential equation can then be written in a form

$$\dot{\Phi} = -a(\Phi - \Phi_g) - (\Phi - \Phi_g) k_\Phi(t),$$

where $k_\Phi(t)$ is the kernel of the memory function and $a$ is a constant. Since the presence of the heavy tail relaxation process is primarily dependent on the ground state, we will approximate that the memory kernel is primarily time- and ground-state dependent ($k_{\Phi_g}(t)$), as the memory effect is absent when the system starts from a non-mixed ground state, despite also proceeding through the mixed states when relaxing. Further, if $k_{\Phi_g}(t) = const$, the equation reduces to a memory-free exponential process $\dot{\Phi} = -a(\Phi - \Phi_g) \Rightarrow \Phi = \Phi_g + const \cdot \exp(-at)$.

Without loss of generality, we can therefore put $a$ to be the constant responsible for the timescale of the fast process and let $k_{\Phi_g}(t)|_{\Phi_g=0} = 0$. If $k_{\Phi_g}(t) = a + C_1/t$, the differential equation directly gives a power law:

$$\dot{\Phi} = -\frac{C_1(\Phi - \Phi_g)}{t};$$

$$\frac{1}{(\Phi - \Phi_g)} d\Phi = \left(-\frac{C_1}{t}\right) dt;$$

$$\ln(\Phi - \Phi_g) = -C_1 \ln(t) + const;$$

$$\Phi = \Phi_g + const \cdot t^{-C_1}.$$

Furthermore, let us assume in the first approximation that the magnitude of the memory process is proportional to the ground state: $k_{\Phi_g}(t) \sim \Phi_g$, and a hyperbolic-type dependence $k_{\Phi_g}(t) \sim t^{-n}$. We can then fit the time dependence of the kernel directly from the data as $k_{\Phi_g}(t) \sim \Phi_g \cdot (t)^{-0.85}$. Taking, to simplify derivation, $k_{\Phi_g}(t) = a + C_2 \Phi_g \cdot (t)^{-0.85}$, the memory term then gives a so-called stretched exponential:

$$\dot{\Phi} = -C_2 \frac{\Phi_g(\Phi - \Phi_g)}{t^{0.85}};$$

$$\ln(\Phi - \Phi_g) = -C_2 \Phi_g t^{0.15} + const;$$

$$\Phi = \Phi_g + C e^{-bt^{0.15}},$$

where $C$ and $b$ are constants. Since this relaxation process is experimentally observed to be orders of magnitude slower than all others, the terms of other processes can be ignored for fitting purposes.

**ACKNOWLEDGMENTS**

The work was supported by U.S. Department of Energy, Office of Science, Office of Basic Energy Sciences, under Contract No. DE-SC0019414 (ultrafast x-ray data collection, analysis, interpretation, and modelling of the behavior O.Yu.G., Z.S., D.W., R.B., and A.S.). This work was supported as part of the Quantum Materials for Energy Efficient Neuromorphic Computing (Q-MEEN-C), an Energy Frontier Research Center funded by the U.S. Department of Energy, Office of Science, Basic Energy Sciences under Award No. DE-SC0019273 (Y.K., A. S., N. H., O. G. S., I.K.S). Use of the Linac Coherent Light Source (LCLS), SLAC National Accelerator Laboratory, is supported by the U.S. Department of Energy, Office of Science, Office of Basic Energy Sciences under Contract No. DE-AC02-76SF00515 (M.S., D.Z.).

**AUTHOR CONTRIBUTIONS**

Initial conceptualization was performed by A.S., I.K.S., and O.G.S. Samples were prepared by Y.K. under supervision of I.K.S. Experiment was performed by O.Yu.G., Z.S., D.W., R.B., A.Sh., N.H., M.S., D.Z., A.S. under supervision of A.S. and O.Yu.G. Initial data processing and visualisation was performed by A. Sh. and N.H. O.Yu.G. performed detailed data analysis, modeling, simulation, and data visualisation. O.Yu.G. wrote the original draft, with review and editing contributions from A.S., I.K.S., and Y.K. and input from all authors.

**COMPETING INTERESTS**

Authors declare no competing interests.

**DATA AVAILABILITY**

Raw data were generated at the Linac Coherent Light Source (LCLS), SLAC National Accelerator Laboratory large-scale facility. Derived data supporting the findings of this study are available from the corresponding author upon request.

# FIGURES

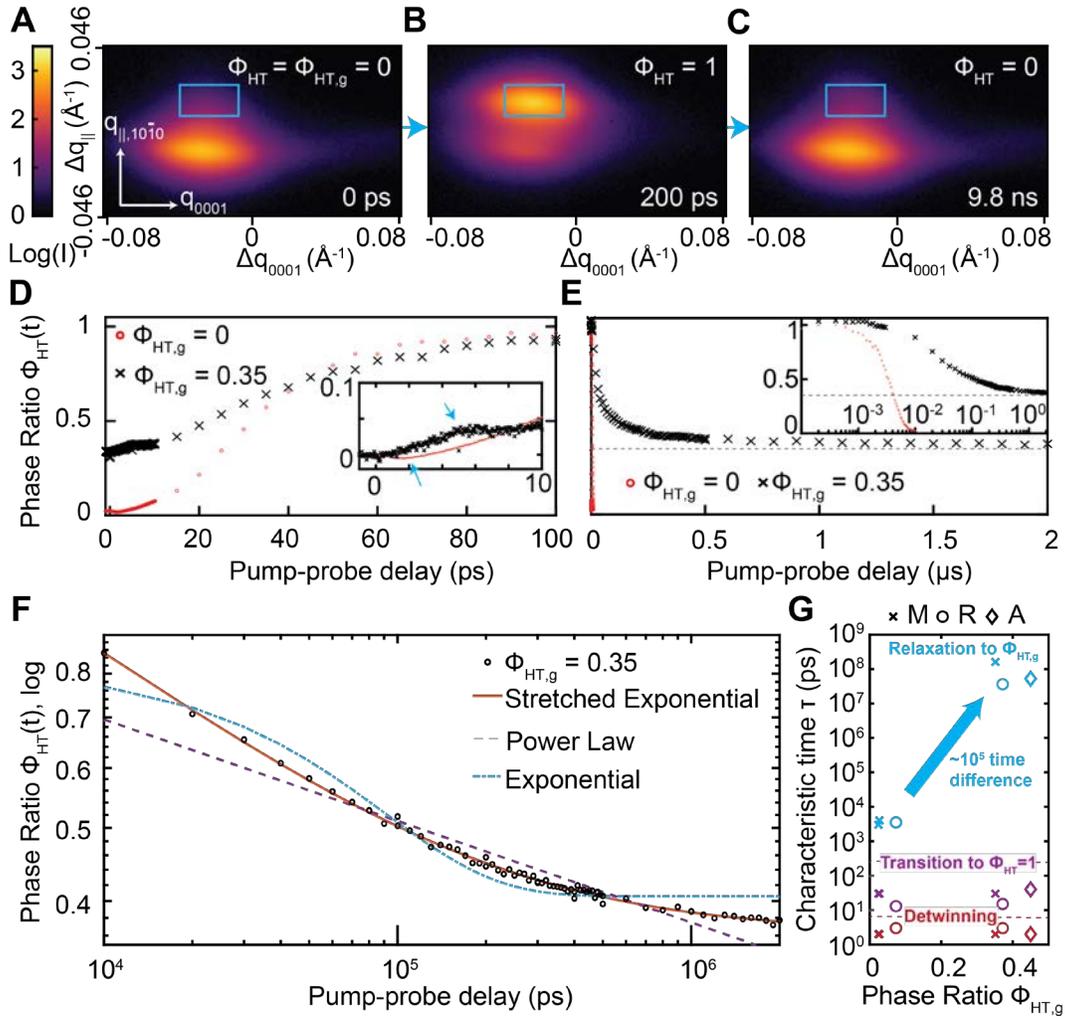

**Fig. 1. Temporal evolution of the phase distribution in a $V_2O_3$ thin film. (A), (B), (C)** Detector images of diffraction patterns around $(30\bar{3}0)$ Bragg peak from the $V_2O_3$ film grown on M-cut $Al_2O_3$. Projected q directions marked on (A). Initial phase ratio $\Phi_{HT,g} = 0$. The images are recorded before photo excitation (A), at $\Phi_{HT,g} = 0$ (B), and after relaxation (C). Intensity is logarithmic. Blue square denotes the region of the diffraction pattern used to calculate the intensity of diffraction from high-temperature peak. **(D)** Phase ratio of the high-temperature phase as a function of pump-probe delay in the first 100 picoseconds after photoexcitation for different pre-excitation phase ratios, $\Phi_{HT,g} = 0$ (red circles) and $\Phi_{HT,g} = 0.35$ (black crests), until saturation at $\Phi_{HT} = 1$. Inset shows relative phase ratio change $\Phi_{HT} - \Phi_{HT,g}$ in the first 10 picoseconds, same markings. Blue arrows denote kinks in the curve corresponding to decrease in monoclinic distortion[7]. **(E)** Phase ratio as a function of delay during relaxation for $\Phi_{HT,g} = 0$ (red circles) and $\Phi_{HT,g} = 0.35$ (black crests), linear (main) and logarithmic (inset) x axis. Dashed line represents $\Phi_{HT,g} = 0.35$. Relaxation for $\Phi_{HT,g} = 0$ is practically instantaneous on the linear scale. **(F)** "Log-log" plot of the relaxation process for $\Phi_{HT,g} = 0.35$ (here black circles), with best fits by stretched exponential, power law, and simple exponential. **(G)** Different characteristic times (detwinning – decrease of monoclinic distortion (red), first stage of the phase transition (magenta), relaxation - second stage of the phase transition (blue)) depending on the ground state and a crystallographic orientation of the film (M - crests, R - circles, A - diamonds). The times are calculated as a second moment (standard deviation) of the distribution (see Materials and Methods).

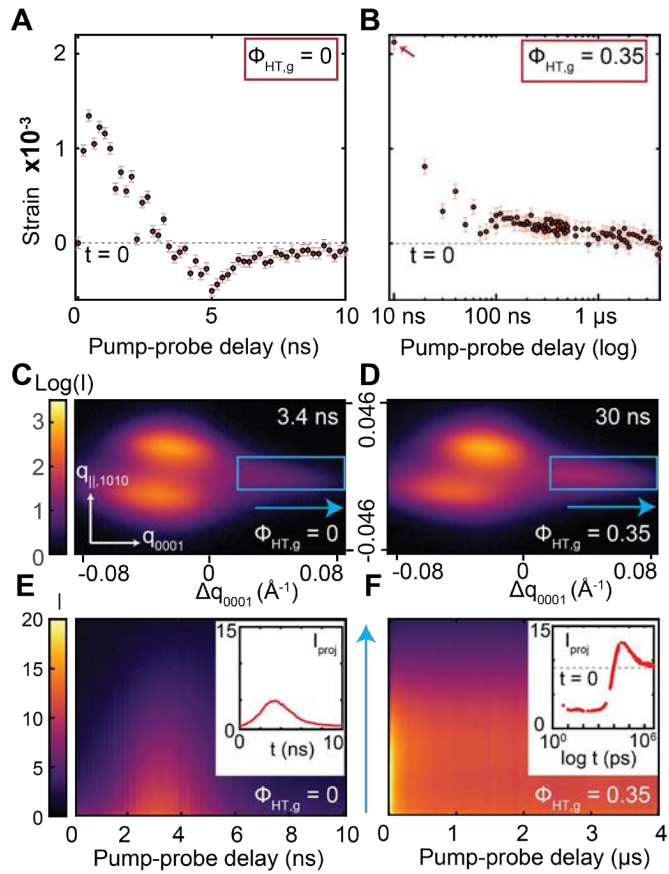

**Fig. 2. Evolution of diffraction peak structure in $V_2O_3$ after photoexcitation. (A), (B)** Strain evolution in the low-temperature peak for $\Phi_{HT,g} = 0$ (A) and $\Phi_{HT,g} = 0.35$ (B) during relaxation. Note log scale for time in (B). Dashed line shows 0. **(C), (D)** Detector images of diffraction patterns around $(30\bar{3}0)$ Bragg peak from the $V_2O_3$ film grown on M-cut $Al_2O_3$ at the time of strongest interface peak. Blue rectangle demonstrates the area of the interface peak to be projected. **(E), (F)** Projected intensity as a function of time during relaxation stage for $\Phi_{HT,g} = 0$ (E) and $\Phi_{HT,g} = 0.35$ (F). Projected area is highlighted by blue rectangle in (C), (D), with arrows representing projection direction. Insets show the evolution of the total intensity in the blue rectangle area. Note logarithmic time scale in the inset of (F).

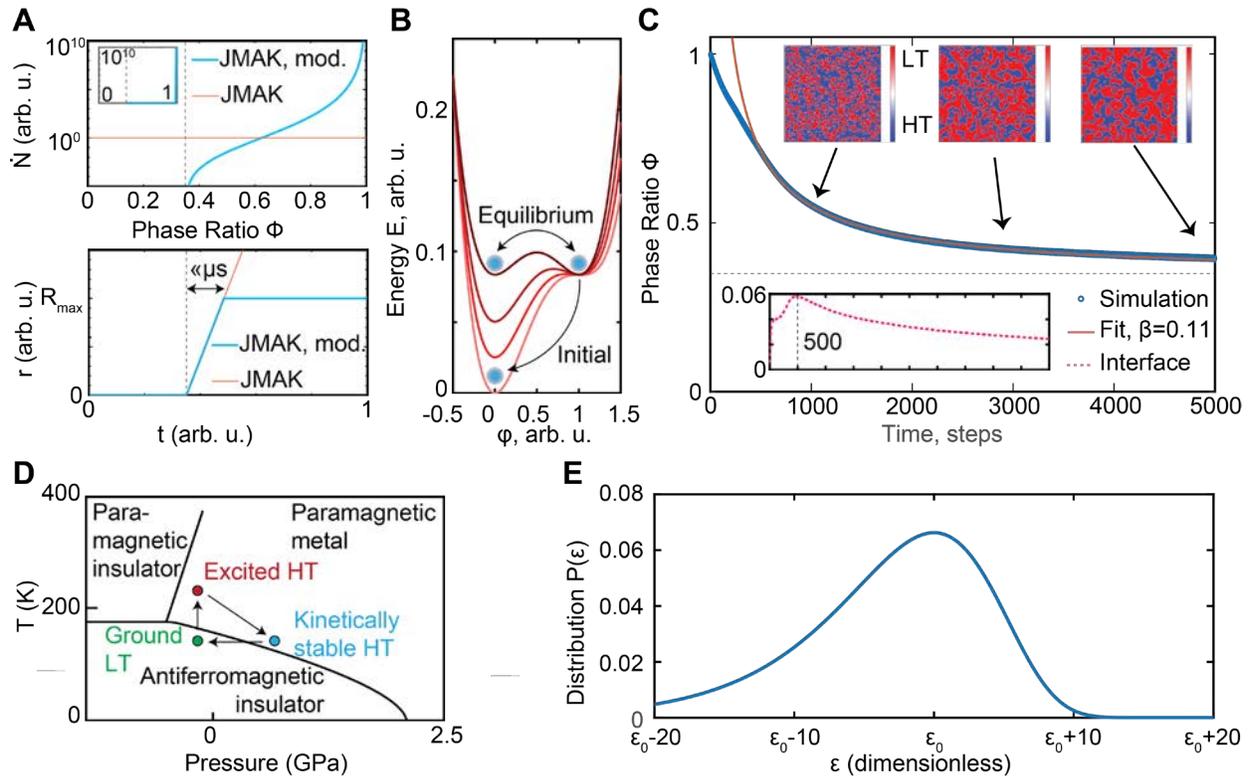

**Fig. 3. Theoretical results and discussion. (A)** Illustration of differences between JMAK and the modified JMAK model nucleation rates and domain radius. Top: nucleation rate (logarithmic scale) as a function of phase ratio: for the classical model (orange, constant) and extracted from experiment assuming modified model (blue). Inset is the same dependence in linear scale, illustrating extreme dependence of nucleation rate for experiment-derived modified JMAK on phase ratio. Bottom: radius of a nucleated domain as a function of time after nucleation. **(B)** Energy "double well" used in Allen-Cahn simulations, with different shades representing the energy surface at different points through relaxation. The system initally easily evolves from 1 – LT phase to 0 – HT phase, but the growing potential barrier and equilibration of wells stalls the transition. **(C)** Results of Allen-Cahn simulations: typical domain distribution (1 is LT phase, 0 is HT phase) and an example of stretched exponential produced (note the exclusion of starting points due to instability kinks). Inset: relative volume of interfaces as a function of simulation time. **(D)** Schematic transition on the phase diagram. **(E)** Barrier height distribution for the fitted stretched exponential in a fully parallel transition.
16

Supplementary Materials for

# Memory in strain-tuned insulator-metal-insulator sequence of transitions after photoexcitation in the Mott material $V_2O_3$

O. Yu. Gorobtsov[1]*, Y. Kalcheim[2,3], Z. Shao[1], A. Shabalin[2], N. Hua[2], D. Weinstock[1]†, R. Bouck[1], M. Seaberg[4], D. Zhu[4], O. G. Shpyrko[2], I. K. Schuller[2], and A. Singer[1]*

Corresponding author: gorobtsov@cornell.edu, asinger@cornell.edu

**Supplementary text**

Additional information on the meaning of scarring

"Scarring", i.e. structure heterogeneities left behind by the pre-transition phase domain mosaic in thin film Mott insulators has been suggested before as a possible memory mechanism during temperature-dependent transport measurements. In our case, scarring would also necessitate signatures of a low-temperature phase left over at high temperatures (after the first stage of the transition, before the relaxation). Instead, the memory effect is present even if the film appears to fully transition into the high-temperature metallic state, but our results currently leave a possibility of low-volume scars left on the domain interfaces.

**Supplementary Figures**

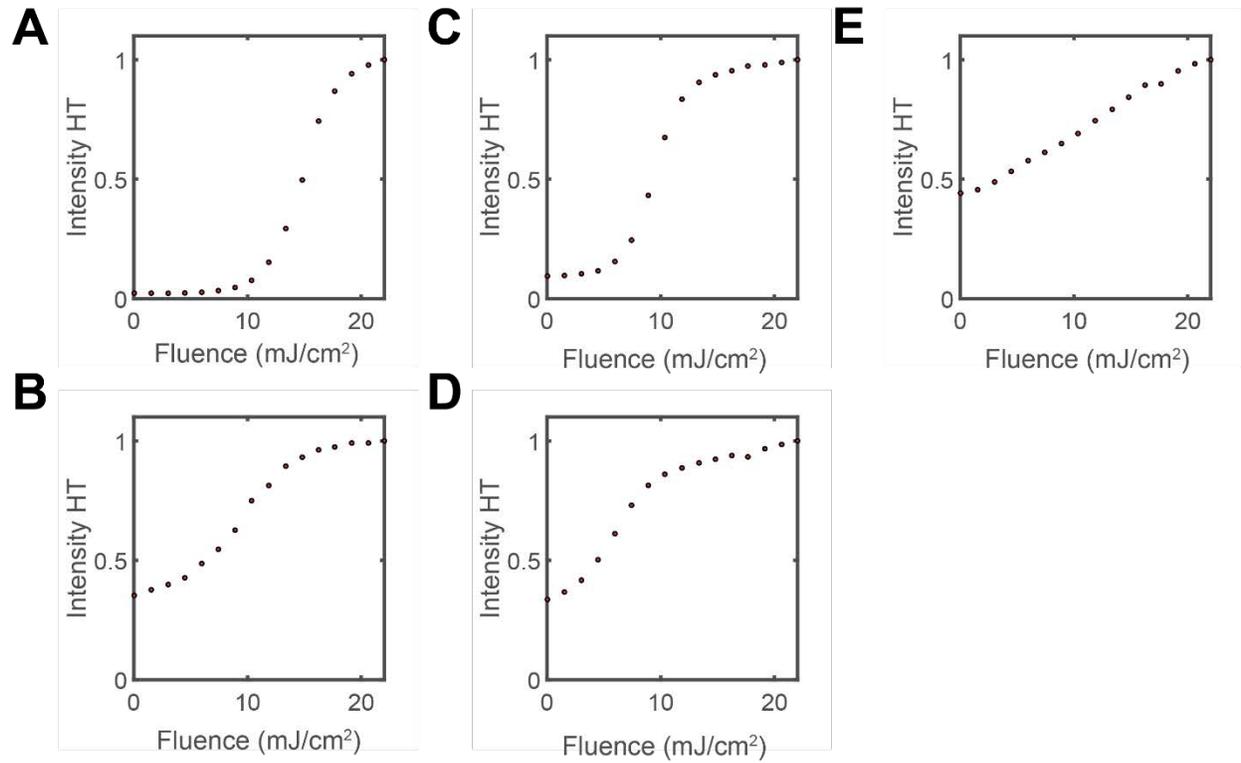

**Fig. S1: Signal saturation of high temperature peak at 100 ps as a function of fluence.** A – M cut, cryojet temperature 110 K, B – M cut, cryojet temperature 140 K, C – R cut, cryojet temperature 100 K, D – R cut, cryojet temperature 140 K, E – A cut, cryojet temperature 100 K. Intensity of the high temperature peak in arbitrary units, normalized to maximum of 1.

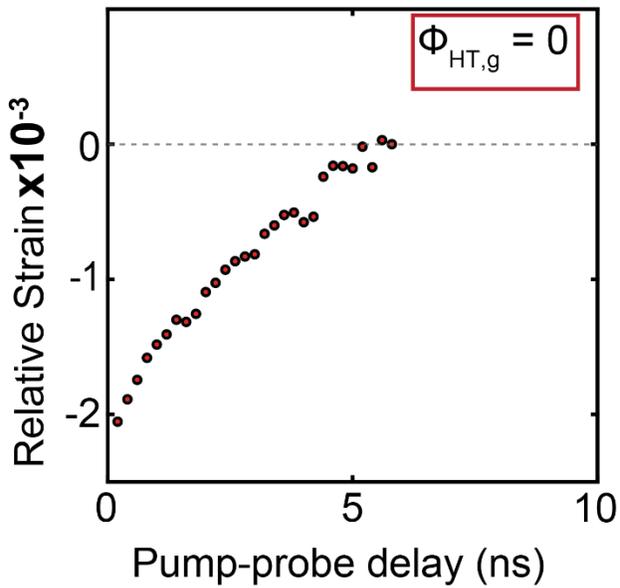

**Fig. S2:** Relative strain evolution in the high-temperature peak for $\Phi_{HT,g} = 0$ during relaxation. Note that difficulties due to intermittent nature of the Bragg peak, its location close to the low temperature Bragg peak, and high potential errors make this particular result insufficiently trustworthy for results, however the direction of evolution of relative strain confirms that the change in lattice constant is due to mechanical effects, not heat transfer.

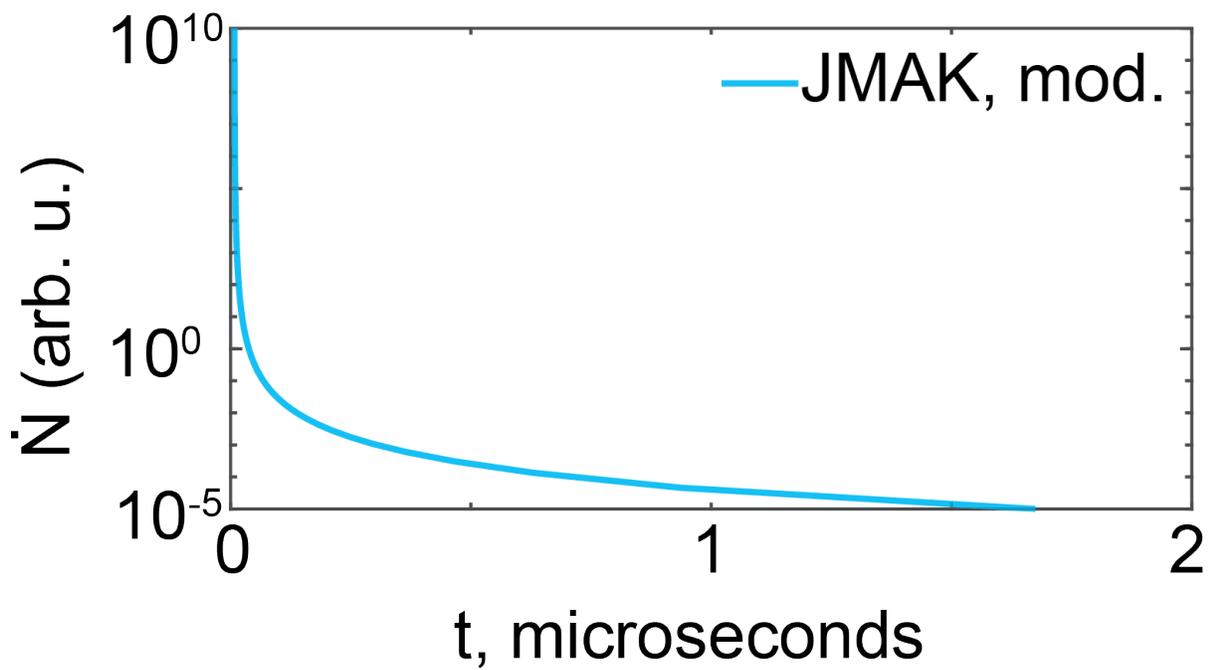

**Fig. S3:** Qualitative dependence of the nucleation rate on time in JMAK model extrapolated from the experiment.